\documentclass[proof]{WileyASNA-v1}
\usepackage{graphicx}
\usepackage{amsmath}
\articletype{Article Type}%

\received{30 April 2018}
\revised{ - }
\accepted{18 June 2018}

\begin{document}

\title{Decomposing blazar spectra into lepto-hadronic emission components}

\author[1,2]{A. Gokus*}
\author[3]{S. Richter}
\author[3]{F. Spanier}
\author[1]{M. Kreter}
\author[1]{M. Kadler}
\author[1]{K. Mannheim}
\author[2]{J. Wilms}

\authormark{A. Gokus \textsc{et al}}

\address[1]{\orgdiv{Institut f\"ur Theoretische Physik und Astrophysik, Universit\"at W\"urzburg}, \orgname{Emil-Fischer-Str. 31}, \orgaddress{\state{97074 W\"urzburg}, \country{Germany}}}

\address[2]{\orgdiv{Dr. Remeis Sternwarte \& ECAP, Universit\"at Erlangen-N\"urnberg}, \orgname{Sternwartstr. 7}, \orgaddress{\state{96049 Bamberg}, \country{Germany}}}

\address[3]{\orgdiv{Center for Space Research}, \orgname{North-West University}, \orgaddress{\state{2520 Potchefstroom}, \country{South Africa}}}

\corres{*A. Gokus \email{andrea.gokus@astro.uni-wuerzburg.de}}

\abstract{Recently reported coincidences between high-energy neutrino events and major blazar outbursts reinforce the relevance of lepto-hadronic emission models for blazars. We study the influence of physical parameters on the neutrino output modeling blazar spectral energy distributions self-consistently assuming a relativistically propagating acceleration zone surrounded by a larger cooling zone. We find that the gross features of the spectral energy distribution can readily be explained with the model. A rigorous test requires time-resolved measurements of blazar spectral energy distributions during an outburst and high-statistics neutrino measurements to discriminate the leptonic and hadronic emission components.}


\jnlcitation{\cname{%
\author{A. Gokus}, 
\author{S. Richter},
\author{F. Spanier}, 
\author{M. Kreter},
\author{M. Kadler},
\author{K. Mannheim}, and
\author{J. Wilms}} (\cyear{2018}), 
\ctitle{Decomposing blazar spectra into lepto-hadronic emission components}, \cjournal{Astronomische Nachrichten}, \cvol{XYZ}.}

\maketitle

\section{Introduction}\label{intro}

\begin{figure}[t]
	\centerline{\includegraphics[width=.48\textwidth]{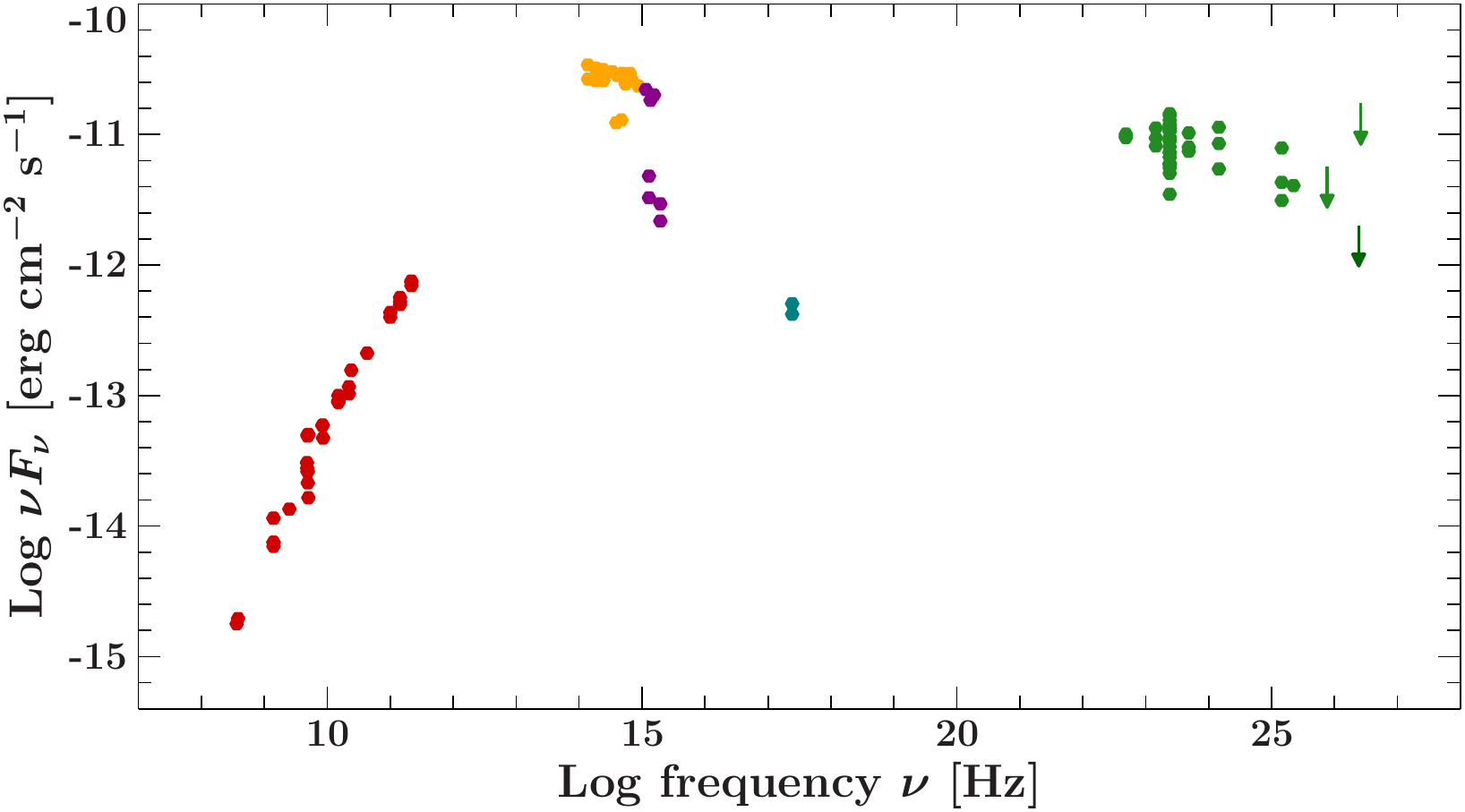}}
	\caption{Historical SED of TXS\,0506+056, using data from the time range 1986 to 2016. Radio observations (red) have been performed by ATCA \citep{atca}, CGRaBS \citep{cgrabs}, CLASSCAT \citep{classcat}, CRATES \citep{crates}, the Green Bank Telescope \citep{gbt1, gbt2, gbt3, gbt4}, the NRAO VLA Sky Survey \citep{nvss}, the Parkes-MIT-NRAO Survey \citep{pmn1, pmn2}, OVRO \citep{ovro1, ovro2}, Planck \citep{planck1, planck2, planck3}, the Texas Interferometer \citep{texas}, VERA \citep{vera}, VLBA \citep{vlba, mojave} and the VLBI Space Observatory Program \citep{vsop1, vsop2}. Optical and near-infrared observations (yellow) have been made by the Kitt Peak National Observatory \citep{kittpeak}, the Swift Satellite \citep{swift} and 2MASS \citep{2mass}. UV observations (purple) were performed by GALEX \citep{galex} and Swift \citep{swift}. Observations in the X-rays (blue) has been done by ROSAT \citep{rosat2,rosat1}. $\gamma$-ray data (green) is from all catalogs from Fermi/LAT \citep{fermi1fgl, fermi2fgl, fermi3fgl} and ARGO2LAC \citep{argo}.\label{fig:archivedata}}
\end{figure}

Active Galactic Nuclei (AGN) with radio jets are promising candidates
for generating high-energy neutrinos with a flux that exceed the atmospheric flux at energies of around 100~TeV \citep{mannheim1995}.
It is commonly adopted that particles are accelerated to relativistic energies in localized regions in the jet, presumably due to shock acceleration, producing non-thermal emission
across the entire observable electromagnetic spectrum. 
The emission from objects that have their jet pointed towards Earth is
boosted by the relativistic bulk motion of the emission regions. These objects are highly time-variable and therefore coined ``blazars'' \citep{angel1980, schlickeiser1996}. 
The spectral energy distribution (SED) of blazars typically exhibits two broad bumps in a log$\nu$-log$\nu$F$_{\nu}$ diagram. The low-energy bump is assumed to result from synchrotron emission by electrons, but the high-energy bump can be explained by either a leptonic or a hadronic scenario \citep[e.g.][]{mannheim1993, dermer1997, sikora2009, boettcher2013}. 
In the leptonic case the X-ray and $\gamma$-ray emission can be produced through Inverse-Compton scattering, where photons scatter off electrons and gain more energy. 
If this process occurs on the synchrotron photons produced by the same population of electrons, it is called synchrotron-self Compton (SSC), otherwise, in the presence of an external photon field, it is external Compton (EC) emission.
If protons are present in the jet, photo-hadronic interactions can take place. The resulting pions from these interactions decay and cause particle cascades that produce $\gamma$-rays, electrons, positrons, muons and also neutrinos \citep{mannheim1989}.
Because neutrinos are of neutral charge and nearly massless, they move undisturbed through space, while other particles and photons created in the jet get deflected by magnetic fields or can be absorbed, respectively.
As both scenarios describe the SEDs of many sources equally well, a distinction could be made for a neutrino detection that can be traced back to a specific AGN.\\
% Neutrino event
The large-volume Cherenkov detector IceCube, which is built in the Antarctic ice at depths from 1450 to 2450\,m, can detect neutrinos in a range from TeV to PeV energies \citep{icecube2006}. Depending on the neutrino flavour, the detected neutrino event has a showerlike or a tracklike appearance for an electron neutrino or a muon neutrino, respectively. While for a cascade-like event the energy of the neutrino can be constrained well, the angular resolution is better for a track-like event.
So far, only a few petaelectron neutrinos have been detected. The neutrino detection of the so-called neutrino "Big Bird" (formally high-energy starting event 35) was a cascade-like event, which had a positional uncertainty of $15.9^{\circ}$.
\citet{kadler2016} showed that the field of interest contained 20 $\gamma$-ray bright AGN, but the blazar PKS B1424$-$418, which showed a major outburst in the months around the detection of the neutrino and it's calorimetric output was sufficiently high to explain the measured neutrino event. 
The recently detected very-high-energy event 170922A is a track-like event that has been spatially and temporally coincident with an increased $\gamma$-ray activity of the blazar TXS\,0506+056, detected by Fermi/LAT \citep{atelfermi}. 
Several multiwavelength follow-up observations confirmed an increase of X-ray \citep{atelswift} and optical activity \citep{atelasas}. MAGIC also detected VHE $\gamma$-rays after the neutrino event \citep{atelmagic}, while H.E.S.S. could not detect any significant $\gamma$-radiation from a point source located in the uncertainty area of the neutrino event \citep{atelhess}.\\
% TXS\,0506+056 source
The blazar TXS\,0506+056 has a redshift of $z=0.3365$ \citep{paiano2018}. The classification of this source is unclear yet as \citet{bllac} identified the source to be a BL Lac according to optical observations, while \citet{crates} state that TXS\,0506+056 is a flat spectrum radio source. The SED of TXS\,0506+056 is shown in Fig.~\ref{fig:archivedata} constructed from archival (non-simultaneous) data. Although the source is present in several surveys and has been observed since 1986, TXS\,0506+056 has not been of special interest so far.

\section{The model}\label{model}
Our goal is to decompose blazar SEDs into the different lepto-hadronic components in order to assess likelihoods for neutrino associations. As an example, we consider the blazar TXS\,0506+056 and model its SED self-consistently, including acceleration and radiation processes \citep{richter2016}. Model parameters were chosen to bracket the observed emission states reasonably well to constrain the range of physical parameters of the emission zone.  However, this approach does not warrant that all possible solutions are already included.  Finding the range of allowed neutrino fluxes from the model is beyond the scope of this paper.
The code by \citet{richter2016} follows a time-dependent procedure with the option for choosing between a purely leptonic approach and a hybrid composition of the jet. The model includes two zones with a spherical geometry, where the acceleration zone with radius $R_{acc}$ is nested within a larger radiation zone with radius $R_{rad}$. 
This approach is similar to a time-independent one zone model. After the injection of a monoenergetic particle population into the acceleration zone, the particles gain energy until an equilibrium between the increasing energy and the synchrotron losses is established for the particles. The acceleration process used here is particle diffusion across a shock front resulting in Fermi-I acceleration.\\
When electrons and protons escape from the acceleration zone, they undergo further radiation and scattering processes, which leads to the production of secondary particles: pions, muons, positrons and neutrinos. These are also treated as time-dependent particle distributions, with the exception of the short-lived pions. The target photon field needed for Inverse Compton scattering and photohadronic interactions is provided by synchrotron photons from electrons, protons, muons, and positrons and the decay of the $\pi^0$.  Bethe-Heitler
pair production was neglected here, because of its inefficiency in the
target spectrum of the test source.
\subsection{The parameter space}
\begin{figure}[t]
	\centerline{\includegraphics[width=.48\textwidth]{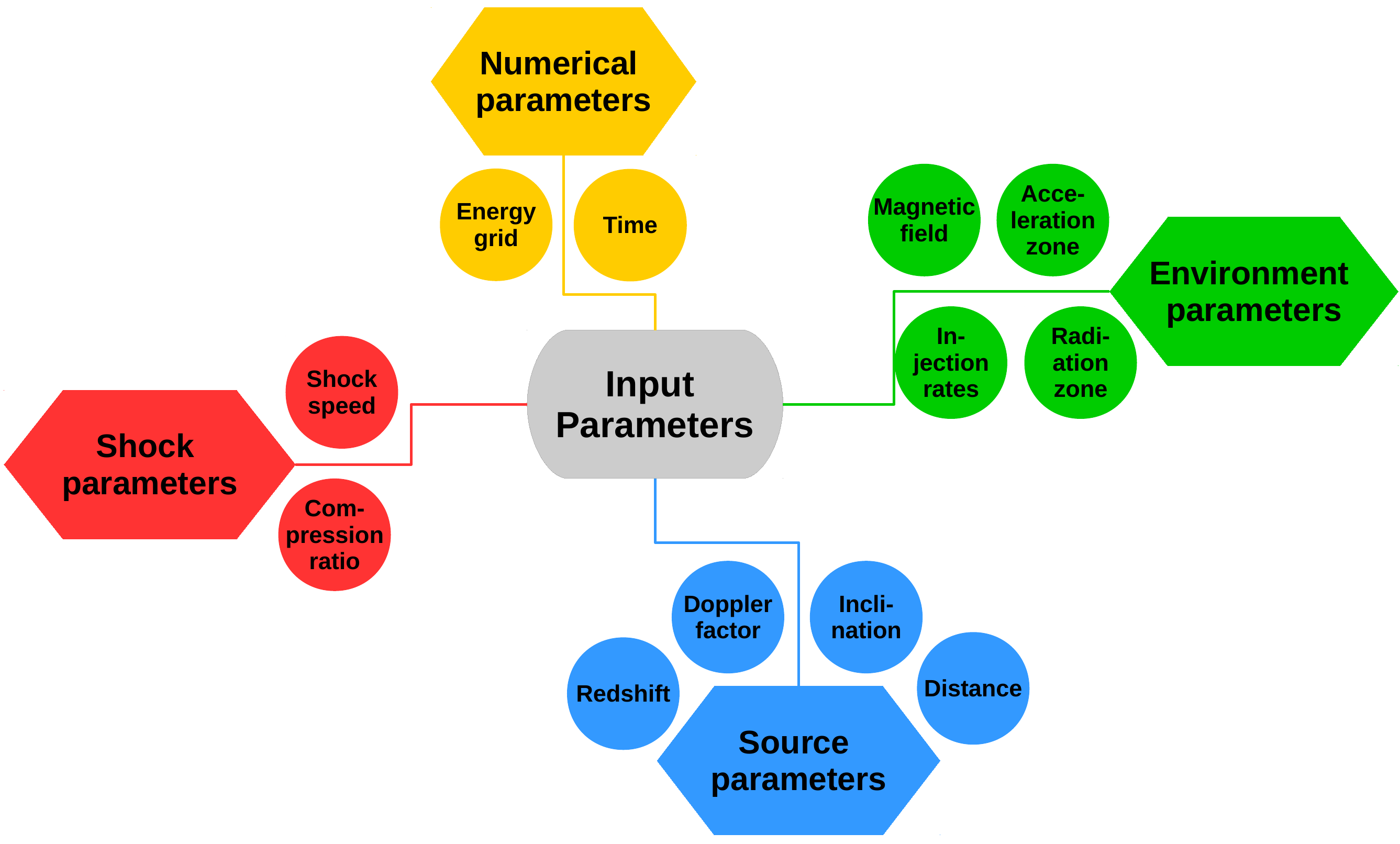}}
	\caption{Schematic overview of the input parameters for our model.\label{fig:parameterspace}}
\end{figure}

The parameter space of the model is large since we take many physical properties into account. An overview of the input parameters is shown in Fig.~\ref{fig:parameterspace}. 
The model is used to calculate a spectrum for all particles involved 
until a steady-state is reached for a given parameter configuration. However, since the calculation is time-dependent and the single time steps need to be short enough to resolve the different processes, the number of time steps to be simulated can range from $10^7$ to $10^{12}$, depending on the parameter configuration. Within practical limitations
set by the available computing power, we varied some crucial parameters, {\it viz.} the size of acceleration- and radiation zone, the strength of the magnetic field, the injection rate of the particles and the Doppler factor. The parameters chosen for TXS\,0506+056 are shown in Table~\ref{tab:parameters}. 
\subsection{Results}
\begin{figure}[]
\centerline{\includegraphics[width=0.48\textwidth]{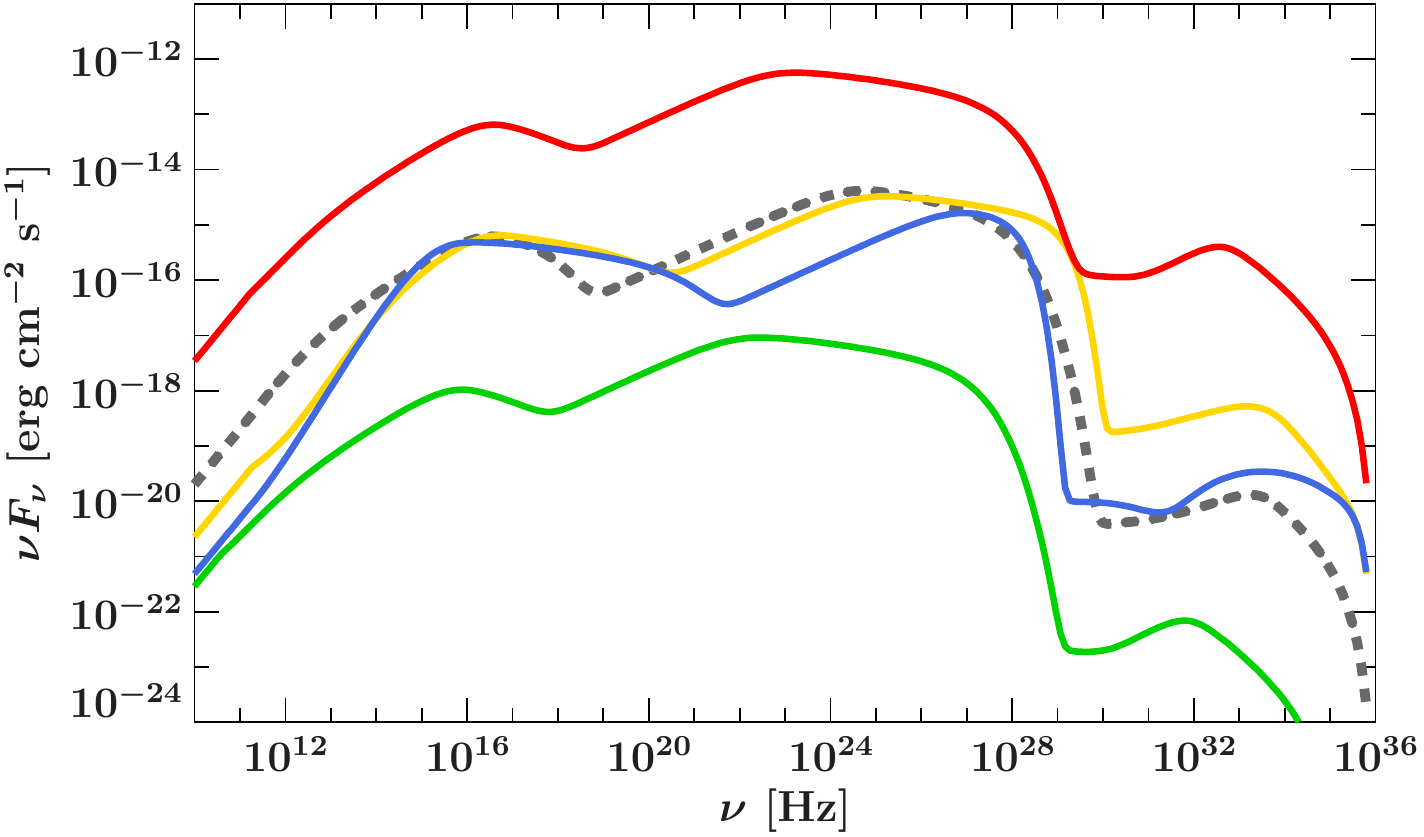}}
	\caption{Different simulated SEDs of TXS\,0506+056 with one varied parameter each. The dotted grey line shows the simulation with the parameter set shown in Table~\ref{tab:parameters}, column 3. The red curve shows an SED for an increased particle rate, while the ratio between protons and electrons has been kept the same. In yellow, a simulation for smaller acceleration and radiation zones is presented. The blue curve displays an SED for a smaller magentic field. In green, the SED was calculated for emission with a lower Doppler factor. \label{fig:allseds}}
\end{figure}

\begin{figure*}[t]
\centerline{\includegraphics[width=0.8\textwidth]{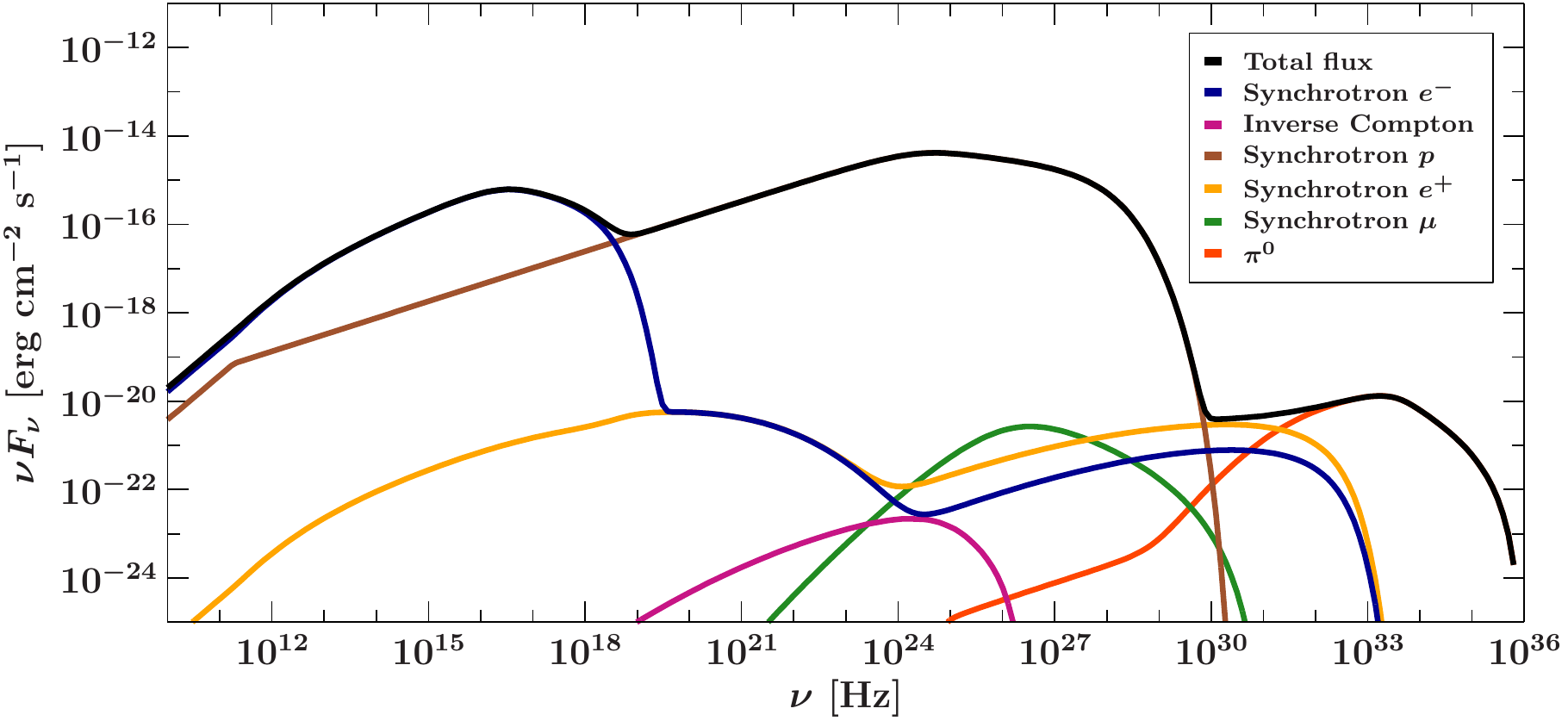}}
	\caption{The simulated SED of TXS\,0506+056 with detailed emission components. The low-energy bump is caused by synchrotron radiation of electrons, as expected, while the high-energy peak is mostly dominated by synchrotron emission of protons.\label{fig:example_SED}}
\end{figure*}

The adopted model parameters used in this work are shown in Table~\ref{tab:parameters}, together with a short description in column 2 and the basic parameter configuration in column 3. The varied parameter values, given in column 4, are used for a parameter study, which is shown in Fig.~\ref{fig:allseds}. One parameter was adopted from the basic parameter set each. The direct comparison can help us understanding how different physical properties influence the overall shape of the SED. All simulations show the typical low- and high-energy bump and an additional third bump with a peak between $10^{32}$ and $10^{34}$ Hz. This component is due to $\gamma$-rays produced by $\pi^0$ particles, which resulted from photohadronic interactions.\\
It is not surprising that the overall flux decreases for a smaller Doppler factor (less beamed flux) and that it increases for a larger particle injection rate. Of special interest is the peak position of the high-energy bump for different parameter values.
The change of the magnetic field strength from $3$ G to $0.3$ G has the biggest influence on the overall shape of the SED. Since synchrotron losses are less important in a weaker magnetic field, this allows electrons with a higher energy to survive.\\
A detailed decomposition of the SED from the basic parameter configuration is given in Fig.~\ref{fig:example_SED}, including the individual components of all radiation processes involved. The low-energy bump, as expected, is caused by synchrotron radiation of electrons. The high-energy bump is a superposition of mostly synchrotron radiation by different particles with the protons contributing the dominant feature. Inverse Compton scattering is present, but strongly suppressed, indicating that photohadronic interactions are favoured in this parameter configuration. The synchrotron emission by positrons, muons and also electrons at higher energies is the result of cascades caused by photohadronic interactions. Protons and $\gamma$-rays scatter and produce either $\pi^0$, which decay into two $\gamma$-rays, or $\pi^{\pm}$, which decay into muons and neutrinos. The high-energetic $\gamma$-rays create electron-positron pairs, while the muons further decay into electrons or positrons, depending on the initial charge of the pion. However, during their lifetime they also produce their own contribution of synchrotron radiation.
\begin{center}
\begin{table*}[t]%
\centering
\caption{Parameters chosen for steady-state simulations of TXS\,0506+056.\label{tab:parameters}}%
\tabcolsep=0pt%
\begin{tabular*}{350pt}{@{\extracolsep\fill}llrr@{\extracolsep\fill}}
\toprule
\textbf{Parameter} & \textbf{Description} & \textbf{Value} & \textbf{Varied value}\tnote{$\dagger$} \\%& \textbf{Colour}\tnote{$\dagger$}  \\
\midrule
$R_{acc}$ & radius of acceleration zone & $3\cdot10^{13}$ cm & $3\cdot 10^{12}$ cm\\% & yellow\\
$R_{rad}$ & radius of radiation zone & $3\cdot10^{16}$ cm & $3\cdot 10^{15}$ cm\\% & yellow\\
$B$ & magnetic field strength & 3 G & $0.3$\text{ G}\\% & blue\\
$\gamma_{inj, \text{e}}$ & injection energy of electrons & $10^4$ & $-$ \\%& $-$ \\
$\gamma_{inj, \text{p}}$ & injection energy of protons & $10$ & $-$ \\%& $-$ \\
$N_{inj, \text{e}}$ & injection rate of electrons & $2\cdot 10^{38}$ s$^{-1}$ & $2\cdot 10^{40}$ s$^{-1}$ \\%& red\\
$N_{inj, \text{p}}$ & injection rate of protons & $8\cdot 10^{40}$ s$^{-1}$ & $8\cdot 10^{42}$ s$^{-1}$ \\%& red\\
$\delta$ & Doppler factor & 25 & 5 \\%& green\\
\bottomrule
\end{tabular*}
\begin{tablenotes}
\item[$\dagger$] Apart from a simulation with the parameters in column 3, four other simulations have been calculated, where for each one parameter has been changed from the original parameter configuration. The changed values for all additional simulations are shown together in column 4. All SEDs are compared in Fig.~\ref{fig:allseds}).
\end{tablenotes}
\end{table*}
\end{center}
Comparing the overall flux from the archival data with the output flux of our model, one can see that we could not find a parameter configuration yet to produce the observed power output from the blazar. Note that here we focused on demonstrating the effects of the various parameters on the model SEDs, instead of showing an optimized fit to the data. This is important, since the existence of a unique best-fit SED is not readily guaranteed for defective data.

\section{Conclusions \& Outlook} 
The long-standing debate about the relative contributions of hadronic versus leptonic emission components in AGN jets is far from being settled, but the detection of neutrinos in the PeV energy regime can help to answer this question as AGN jets are one of the most probable accelerators to produce particles to sufficiently high energies. The self-consistent model employed here, simultaneously treats synchrotron-self Compton and photohadronic processes together with particle acceleration to study the blazar SED.
We find that several parameters can change the shape and intensity of the SED, leading to very different expected neutrino fluxes. Until high-statistics measurements will become available, time-resolved studies of blazar spectral energy distributions 
are a promising avenue to resolve the issue in the frame of the model assumptions (Fermi acceleration at relativistically moving shocks), emphasizing the need for coordinated multi-frequency campaigns for blazar observations.
\section*{Acknowledgments}
This research has been partially funded by the Bundesministerium f\"ur Wirtschaft und Technologie under Deutsches Zentrum f\"ur Luft- und Raumfahrt grant number 50OR1607, and by the
Bayerisch-Tschechische Hochschulagentur (BTHA)
under grant number BTHA-AP-2018-18.
This research has made use of a collection of ISIS functions (ISISscripts) provided by ECAP/Remeis observatory and MIT (http://www.sternwarte.uni-erlangen.de/isis/). 
Part of this work is based on archival data, software or online services provided by the Space Science Data Center - ASI. 
This research has made use of the NASA/IPAC Extragalactic Database (NED) which is operated by the Jet Propulsion Laboratory, California Institute of Technology, under contract with the National Aeronautics and Space Administration. 
%\section*{Author Biography}
%\begin{biography}{\includegraphics[width=60pt,height=70pt,draft]{portrait.eps}}{\textbf{Andrea Gokus.} Andrea Gokus received her Bachelor's and Master's Degree in Physics at the Friedrich-Alexander University of Erlangen-Nuremberg. She is currently working as a PhD student at the institute for theoretical physics and astrophysics at the University of Wuerzburg and the Dr. Karl Remeis Observatory Bamberg.}
%\end{biography}
\bibliography{Wiley-ASNA}

\end{document}